\documentclass[preprint]{ptephy_om}
\preprintnumber{RIKEN-iTHEMS-Report-25, KYUSHU-HET-348}

\usepackage{bm}
\usepackage{booktabs}

\numberwithin{equation}{section}

\usepackage{graphicx,xcolor,tikz}
\usetikzlibrary{decorations.pathmorphing}
\tikzset{snake it/.style={decorate, decoration=snake}}

\usepackage{hyperref}
\usepackage{orcidlink}

\newcommand{\revone}[1]{#1}

\newcommand{\revtwo}[1]{#1}

\newcommand{\revthree}[1]{#1}

\title{Geometric phase \revone{from encircling an} exceptional point \revone{of a} quantum resonance in \revone{the} complex-scaling method\thanks{This work is dedicated to Takuma Matsumoto and Yushin Yamada.}}

\author[1]{Okuto Morikawa \orcidlink{0000-0002-0044-4491}}
\affil[1]{Center for Interdisciplinary Theoretical and Mathematical Sciences (iTHEMS),
RIKEN, Wako 351-0198, Japan
\email{okuto.morikawa@riken.jp}}

\author[2]{Shoya Ogawa \orcidlink{0000-0003-0900-2486}}
\affil[2]{Department of Physics, Kyushu University, 744 Motooka, Nishi-ku,
Fukuoka 819-0395, Japan}

\author[2]{Soma Onoda \orcidlink{0009-0008-9886-9279}}

\begin{document}
\begin{abstract}
\revthree{Non-Hermitian operators are now routinely used to describe few-mode systems such as optical resonators and superconducting qubits, and exceptional points (EPs) are defective spectral singularities of such non-Hermitian operators. In contrast, the scattering-theoretic formulation of EP physics for unbounded Hamiltonians remains less settled. In this work, we formulate the geometric phase associated with encircling an EP when the underlying eigenstates are quantum resonances within a one-dimensional scattering model.}
To do this, we employ the complex-scaling method, where resonance poles of the S matrix are realized as discrete eigenvalues of the non-Hermitian dilated Hamiltonian, to construct situations in which resonant and scattering states coalesce into an EP in the complex energy plane, that is, the resonance pole is embedded into the continuum spectrum. We analyze the self-orthogonality in the vicinity of an EP, the Berry phase, and the Chern characteristic.
\revthree{Our results clarify how EP branch structure and geometric holonomy arise directly from resonance poles in scattering theory, thereby connecting non-Hermitian spectral topology with the traditional theory of quantum resonances.}
\end{abstract}
\maketitle

\tableofcontents

\section{Introduction}
\revone{Non-Hermitian operators have become a central language for describing wave and quantum systems with gain, loss, or openness, and exceptional points (EPs) \revthree{are} 
\revtwo{\revthree{defective} spectral singularities that play a central role in non-Hermitian spectral topology}~\cite{Ashida2020NonHermitian}. 
While \revtwo{non-Hermitian generators appear either as reduced descriptions or as operationally realized evolutions}, an equally important arena is 
\revthree{quantum scattering theory}, where resonances appear as poles of the S-matrix and where the relevant spectral singularities are tied to the analytic structure of these poles.}
\revthree{Here, by ``quantum scattering theory'' we mean the resonance problem for an unbounded Hamiltonian, as opposed to finite-dimensional effective Hamiltonians obtained, for example, from mode truncation, Feshbach projection, or phenomenological descriptions of open-system dynamics. Our aim is therefore not to characterize non-Hermitian physics in general as merely effective, but to identify a concrete resonance-based mechanism by which EP-related branch structure emerges directly in such a scattering framework.}

\subsection{Overview}
Over the past two decades, non-Hermitian operators have emerged as a central concept in various areas of wave and quantum physics. In optics, mechanics, electrical circuits, acoustics, and related platforms, gain, loss, and nonreciprocal couplings are routinely engineered and described in terms of \revtwo{non-Hermitian Hamiltonians which are not only useful as reduced descriptions in resonance and postselected settings, but can also be realized operationally in experiments~\cite{Foroozani:2015jfz,Naghiloo:2019dqw}}. These systems have enabled the realization of parity-time (PT) symmetry, exceptional points (EPs), and non-Hermitian topological phases. They exhibit phenomena such as unidirectional transparency, coherent perfect absorption, anomalous bulk-boundary correspondence, and enhanced sensitivity near EPs~\cite{Ashida2020NonHermitian,Gong2018Topological,Yao2018EdgeStates}. In \revone{both few-mode and extended} settings, \revtwo{non-Hermitian Hamiltonians are by now well established as useful models of wave and quantum phenomena}.

\revone{More broadly,} the foundations of non-Hermitian quantum mechanics in \textit{genuinely infinite-dimensional} \revone{high-energy} settings---quantum scattering theory, unbounded Hamiltonian operators, and quantum field theory (QFT)---remain much less settled. From the mathematical side, a variety of rigorous tools exist: Gamow (Siegert) vectors and rigged Hilbert spaces for resonances, the complex-scaling method (CSM) and related analytic dilations, and Feshbach projection  \revtwo{leading to reduced non-Hermitian operators}~\footnote{%
\revtwo{Non-Hermiticity can qualitatively alter spectral topology and spatial asymptotics, as seen in non-Bloch band theory and the non-Hermitian skin effect~\cite{Yao2018EdgeStates,Gong2018Topological,Hetenyi:2025a,Hetenyi:2025b,Alase:2016tqv}. This motivates a careful treatment of scattering and resonance phenomena beyond standard Hermitian settings.}
\revtwo{Related non-Hermitian operators also appear, for example, in Feshbach-type treatments of resonance/scattering problems and as no-jump generators in quantum-trajectory formulations; see, e.g., Refs.~\cite{Gneiting:2020pew,Piilo:2008dmi}.}}
\cite{Moiseyev2011NHQM,Ashida2020NonHermitian}. However, the connection between this older functional-analytic machinery and the newer language of non-Hermitian physics---EPs, non-Hermitian topology, non-Hermitian quantum speed limits, and so on---is only partially understood. In particular, the precise conditions under which non-Hermitian operators provide a faithful and controllable description of scattering resonances, rather than merely a numerical tool, are still being clarified.\footnote{A closely related issue is the time evolution of unstable quantum states. It has been known that the decay of an unstable state is generically nonexponential (often power law), both in nonrelativistic quantum mechanics and in QFT~\cite{GarciaCalderon2017NonExp,Giacosa2012NonExpQFTQM}. More recently, it has been emphasized that nonexponential decay can be significantly enhanced near EPs and spectral thresholds in continuum models; this leads to decay laws that differ qualitatively from those predicted by Fermi's golden rule~\cite{Garmon2017NonExpThreshold,Garmon2021AnomalousEP}.}

\subsection{Non-Hermitian physics for scattering resonance phenomena}

\revtwo{This work studies how scattering resonances give rise to EP-related branch structure and geometric holonomy in a non-Hermitian description provided by the complex-scaling method.} Here, analytically tractable one-dimensional scattering models provide an invaluable testing ground. Also, we employ CSM where the resonance pole possesses the resonant state as a normalizable wave function; it is a nonunitary similarity transformation and generates a non-Hermitian quantum mechanics with resonance as pseudobound states. The functional-analytic approach of quantum mechanics under CSM naturally leads to a rigged Hilbert space for resonance and continuum scattering~\cite{MorikawaOgawa2025JHEP,MorikawaOgawa2025Continuum}. The number of the regularized resonant states in the corresponding Hilbert space depends on the scaling angle, and hence changing this angle should encounter non-Hermitian EPs such that a resonance pole is embedded in the complex-scaled scattering spectrum. This setting provides a simple but nontrivial example of EP physics in infinite-dimensional quantum scattering theory.
\revone{Our EP is not introduced as a formal analogy but as the branch-point singularity of a resonance pole in a scattering problem, which directly controls the monodromy and holonomy of the resonance eigenstate under parameter cycles.}

By enlarging the parameter space of the potential under CSM, in this paper, we will construct situations in which a resonance pole and a continuum state coalesce into an EP in the complex energy plane and examine the behavior of the associated eigenfunctions. This allows us to ask, within a fully specified one-dimensional model, what it really means to cross a resonance or to pass through an EP from the standpoint of scattering theory and non-Hermitian spectral theory.
\revone{Complex scaling allows us to treat resonances as discrete eigenstates and to compute their geometric response; the relevant EP is the spectral singularity where a resonance pole meets the rotated continuum, providing a concrete scattering-theoretic mechanism for holonomy.}
Then, we reproduce many aspects of non-Hermitian physics: self-orthogonality (non-diagonalizable Hamiltonian), geometric Berry phase, and (a renormalization theory of) Chern-characteristic/holonomy class.

\revtwo{To analyze the resonance-continuum coalescence in a form that can be followed explicitly, we use a momentum-bin discretization of the complex-scaled continuum. The resulting binned states are normalized by a Kronecker delta, which makes the relevant formulas and branch structure more transparent. By contrast, the original continuum states are Dirac-delta normalized, so the corresponding expressions are distributional; this is not merely technical, but will later be related, from a renormalization-theoretic viewpoint, to the Chern-class structure. Technical details and notation are collected in Appendix~\ref{sec:bin_method}.}

\revone{We remark on the physical meaning of the parameter loop and possible experimental signatures. In a scattering setting, adiabatically varying a control parameter should be understood in a quasi-static sense: one considers a slowly time-dependent potential $V(x,t)=V[x;\lambda(t)]$ such that an instantaneous stationary scattering description (and hence an instantaneous S matrix) is well defined at each time. A natural criterion is that the driving timescale $T_{\rm drive}$ is much longer than the intrinsic scattering response time, which can be estimated by the Wigner time delay $\tau_W(E;\lambda)$, i.e.\ $T_{\rm drive}\gg\tau_W(E;\lambda)$.
Under such a slow closed cycle in parameter space enclosing the branch point, the holonomy discussed in this work is expected to be encoded at the level of the S matrix. Therefore, it may leave a trace in a scattering phase shift. A detailed quantitative modeling is beyond the scope of the present work and is left for future investigation.}
We hope that these results inform the broader question of when and how non-Hermitian Hamiltonians provide a consistent and physically meaningful description of scattering and decay in \revone{general} quantum systems \revone{including QFTs}.

\subsection{Remark on terminology}
\revone{Throughout this paper, we use ``EP'' to mean a defective spectral singularity (coalescence of eigenvalues and eigenvectors) of a non-self-adjoint operator.
In our scattering setting formulated via complex scaling, such defectiveness occurs when a resonance pole reaches the complex-rotated continuum, producing a branch point in parameter space. Thus, the EP studied here is a resonance-physics realization of the general EP notion in an infinite-dimensional operator.}

\revone{It is however quite difficult to treat the continuum spectrum technically. Introducing a discretization scheme provides a more tractable setting.
Then, an EP is defined as a parameter value at which two (or more) eigenvalues and eigenvectors are degenerate and the operator becomes defective.
}
\revtwo{In the present work, we focus on the branch-point structure that arises in resonance physics formulated through the complex-scaling method and momentum-binned continuum spectrum.}
A kind of continuum limit will be considered in the context of renormalization group theory.

\revone{Resonance EP under the CSM} will be equivalently characterized by a branch-point singularity in the analytic continuation of the resonance pole as a function of an additional parameter $\lambda$ in the theory.
Depending on which parameter is varied, the same phenomenon can be discussed either as a branch point in the $\lambda$ plane (at fixed $\theta$) or as a critical value of the rotation angle $\theta$ (at fixed $\lambda$).
Because the eigenstate is intrinsically multivalued around such a branch point, geometric quantities (holonomy or Berry phase)
are naturally defined on a branched cover of the punctured $\lambda$ plane; correspondingly, a naive Chern-type invariant
defined on a single sheet may appear fractional and becomes integer valued only after passing to the appropriate cover.

\revtwo{In physical terms, the parameter loop of~$\lambda$ considered here should be understood as a quasistatic cycle of a controllable scattering potential. The significance of the present analysis is that the branch-point holonomy of the resonance sector can then, at least in principle, be encoded in the corresponding slowly varying scattering response, rather than remaining a purely formal feature of the discretized representation.}

\section{Resonance wave function in complex-scaling method}
\subsection{Complex scaling and the ABC theorem}
\revone{Resonances in scattering correspond to poles of the S matrix at complex energies
\begin{equation}
E = E_R - \frac{i}{2}\Gamma,
\end{equation}
where $E_{R}$ is a resonant energy, and $\Gamma>0$ is a decay width.
A resonance wave function is characterized by the purely outgoing (Siegert) boundary condition~\cite{Siegert:1939}. 
In one-dimensional scattering, the outgoing solutions behave asymptotically as
\begin{equation}
\psi(x)\sim
\begin{cases}
e^{+ikx} & x\to +\infty,\\
e^{-ikx} & x\to -\infty,
\end{cases}
\end{equation}
with a complex wave number $k=\sqrt{2mE}/\hbar$.
Because $\mathrm{Im}\,k\neq 0$ in general, such solutions diverge exponentially in at least one direction and are not square integrable,
\begin{equation}
\psi \notin L^2(\mathbb{R}).
\end{equation}
Standard methods determine reflection and transmission for real $E$, but do not by themselves turn resonance poles into a discrete, well-posed eigenvalue problem.}

\revone{The complex-scaling method (CSM) provides a framework that exposes resonance poles as discrete eigenvalues. 
One performs the complex coordinate rotation
\begin{equation}
x \mapsto x e^{i\theta},\qquad 0<\theta<\frac{\pi}{4},
\end{equation}
implemented by a nonunitary transformation $U_\theta$,
which maps the Hamiltonian $H$ to the complex-scaled operator
\begin{equation}
H_\theta = U_\theta H U_\theta^{-1}.
\end{equation}
Operationally, complex scaling has two key consequences: (i) purely outgoing resonance solutions, which are non-$L^2$ on the real axis, become square integrable after scaling, and (ii) resonances can be computed as discrete eigenpairs $(E,\psi_\theta)$ of $H_\theta$, rather than being inferred indirectly from peaks or rapid phase variations on the real energy axis.}

\revone{The mathematical basis for these statements is provided by the Aguilar--Balslev--Combes (ABC) theorem~\cite{Aguilar:1971ve,Balslev:1971vb}, 
which implies the following properties of $H_\theta$:
\begin{enumerate}
    \item \textit{Square integrability of resonances.} Resonant solutions are represented by $L^2$ eigenfunctions of $H_\theta$, similarly to bound states.
    \item \textit{$\theta$ stability.} Bound-state energies and resonance eigenvalues are invariant under variations of the scaling angle $\theta$ within an analyticity sector.
    \item \textit{Rotation of the continuum.} The continuous spectrum, starting from physical threshold energies, is rotated clockwise by an angle $2\theta$ from the positive real axis in the complex energy plane.
\end{enumerate}
Thus, at finite $\theta$ the resonance problem is reduced to a discrete (non-Hermitian) eigenvalue problem for $H_\theta$.}

\subsection{Solving method within complex-scaling method}
Our considerations in this paper will be applicable quite generically to scattering theory in one dimension. To illustrate it, we show explicit computations in an exactly solvable model: the inverted Rosen--Morse potential or P\"oschl--Teller potential. It is well known that this system supports barrier resonance.

For the inverted Rosen--Morse potential, the Schr\"{o}dinger equation \revone{under the CSM} is given by
\begin{align}
    \left[-\frac{\hbar^2}{2m} \frac{d^2}{dx'^2} + \frac{\lambda}{\cosh^2 \beta x'}\right] \psi(x')
    = E \psi(x') ,
    \label{eq:CSM-eq}
\end{align}
where we have replaced $x\in\mathbb{R}$ by $x'=xe^{i\theta}\in\mathbb{C}$ with the scaling angle $\theta$ and assume $\lambda>0$.
\revone{$m$ is the mass parameter, $\lambda$ and $\beta$ determine the height and the width of the potential, respectively.}
\revone{For simplicity---namely, to identify the differential equation with a hypergeometric-type equation---we introduce several somewhat technical but convenient variables.}
\revone{Letting $\xi=\tanh\beta x'$}, the equation is represented as
\begin{align}
    \left[ \frac{d}{d\xi} (1-\xi^2)\frac{d}{d\xi} +s(s+1) - \frac{\kappa^2}{1-\xi^2} \right] \psi(x')
    = 0 ,\\
    \kappa = \frac{\sqrt{-2mE}}{\beta\hbar},
    \qquad
    s = \frac{1}{2} \left(-1 + \sqrt{1-\frac{8m\lambda}{\beta^2\hbar^2}}\right).
    \label{eq:CSM-reduced-eq}
\end{align}
Substituting $\psi(x')=(1-\xi^2)^{\frac{\kappa}{2}}\omega(\xi)$ into Eq.~\eqref{eq:CSM-reduced-eq}, we obtain the hypergeometric differential equation \revone{explicitly in terms of~$\beta$, $\kappa$, $s$, and $u$}
\begin{align}
    \left[
    u(1-u)\frac{d^2}{d\xi^2} + (\kappa + 1) (1 - 2 u) \frac{d}{d\xi} - (\kappa - s) (\kappa + s + 1) \right] \omega(\xi) = 0, \quad u=\frac{1-\xi}{2} .
\end{align}
The solution of this equation, which is regular at $x=0$, is given by
\begin{align}
    \psi_{\mathrm{CSM}}(x)= (1-\xi^2)^{\frac{\kappa}{2}} 
    F\left( \kappa-s, \kappa+s+1, \kappa+1, \frac{1-\xi}{2} \right).
\end{align}
Here, $F$ is the Gauss hypergeometric function.
In general, as $x\to-\infty$ ($\xi\to-1$), the Gauss hypergeometric function diverges.
This naive wave function is missing in Hilbert space but is called the Gamow state in a physical sense.
However, physical wave functions of resonant states under CSM are normalizable because of the ABC theorem.
To ensure the normalizability of the wave function, it is well-known that the hypergeometric function should be expressed as a finite-order polynomial.
This requirement leads to the condition, $\kappa-s=-n$ ($n\in\mathbb{Z}_{\geq0}$), which gives the exact energy spectrum of resonance,
\begin{align}
    E^{\mathrm{R}}_n = \frac{\hbar^2\beta^2}{8m} \left[\sqrt{\frac{8m\lambda}{\beta^2 \hbar^2} - 1} - i(2n + 1) \right]^2.
\end{align}

\revone{We have introduced some somewhat technical but convenient variables. In what follows, the resonance solutions obtained via the CSM, $\psi_{\mathrm{CSM}}$, (given by hypergeometric functions times an overall factor) and the discrete resonance energy spectrum, $E^{\mathrm{R}}_n$, (lying in the fourth quadrant) will be of primary importance.}

\subsection{Analyticity of~$\lambda$ near resonance pole}

\begin{figure}[t]
    \centering
\begin{tikzpicture}
    \draw[->,thick] (-1,0) -- (5,0) node[right] {$\re E$};
    \draw[->,thick] (0,-5) -- (0,1) node[above] {$\im E$};
    \draw[dashed,very thick] (0,0) -- (5,-8/3) node[above] {$\theta_n$};
    \draw[dashed,very thick] (0,0) -- (35/24,-5) node[right] {$\theta_{n+1}$};
    \draw[ultra thick,blue] (0,0) -- (4,-5) node[above] {$\theta$};
    \draw[very thick,blue,<->] (1,0) arc(0:-50:1) node[midway,right] {$2\theta$};
    \fill[red] (5/2,-4/3) circle(5pt) node[below left] {$E^{\mathrm{R}}_n$};
    \draw[red,ultra thick] (7/6,-4) circle(5pt) node[above right] {$E^{\mathrm{R}}_{n+1}$};
    \draw[dotted,teal,ultra thick] (17/6,0) arc(0:-25:17/6) node[midway,right] {resonant state at $E^{\mathrm{R}}_{0,1,\dots,n}$};
    \draw[dotted,teal,ultra thick] (0,-25/6) arc(270:285:25/6) node[midway,below] {pole};
    \draw[<-,teal,ultra thick] (16/5,-4) -- (4,-4) node[right] {conti. state};
\end{tikzpicture}
    \caption{Schematic energy-plane picture of complex scaling. The complex-scaled continuum spectrum is rotated by $-2\theta$ (blue) from the positive real axis. We set $\theta_n<\theta<\theta_{n+1}$. Resonance poles at $E_{n+1,n+2\dots}^{\mathrm{R}}$ appear below the rotated cut, and Gamow states at $E_{0,1,\dots,n}^{\mathrm{R}}$ become regularized as discrete states.}
    \label{fig:theta_on_E}
\end{figure}
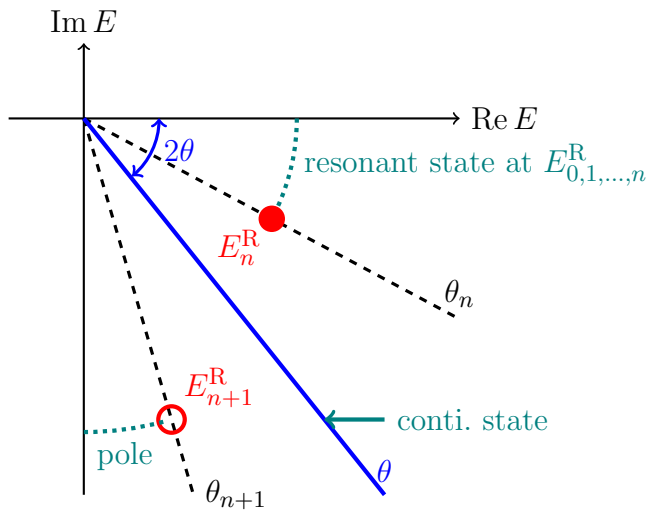

The ABC theorem states that, \revone{as mentioned above}, the continuum spectrum is rotated by~$-2\theta$, $E_{\mathrm{CSM}}(\theta) = E e^{-2i\theta}$ with the positive and continuum energy~$E$ of an incoming wave as the parameter in scattering problem.
Let $\theta_n$ be an angle for which the resonance pole with~$E^{\mathrm{R}}_n$ contacts with the complex-scaled continuum spectrum, i.e., there exists $E$ such that $E^{\mathrm{R}}_n = E_{\mathrm{CSM}}(\theta_n)$.
The explicit form of~$\theta_n$ would be given by
\begin{align}
    \theta_n = \frac{1}{2} \arctan\left(- \frac{\im E^{\mathrm{R}}_n}{\re E^{\mathrm{R}}_n}\right) .
\end{align}
The number of basis vectors (wave functions) spanning the rigged Hilbert space is the total number of the bound states and continuum states when $0\leq\theta<\theta_0$, while if we assume that $\theta_n<\theta<\theta_{n+1}$ then the number of bases is the original number of the bound and continuum states plus the number of the resonant states regularized by CSM, $n+1$, see Fig.~\ref{fig:theta_on_E}. In other words, $\theta=\theta_n$ is an exceptional point (EP) where the number of (normalizable) wave functions changes (we will define more precisely).

\begin{figure}[t]
    \centering
    \includegraphics[width=0.5\columnwidth]{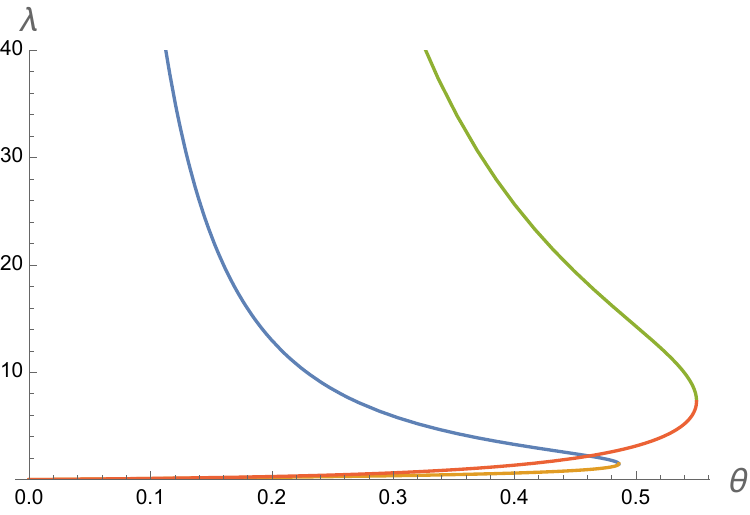}
    \caption{$\lambda_{0,1}^\pm$ as functions of~$\theta$. $\frac{\beta^2\hbar^2}{4m}=1$. $\lambda$ should be above the blue curve which denotes $\lambda_0^+$ ($\lambda$ should be larger than it), or be below the orange curve which is $\lambda_0^-$ ($\lambda$ should be smaller than it). The green and red curves correspond to~$\lambda_1^+$ and~$\lambda_1^-$, respectively, and hence $\lambda$ should be inside the region between these curves. Eventually, we find that $\lambda$ is larger than~$\lambda_0^+$ and smaller than~$\lambda_1^+$ as long as $\theta$ is not too large.}
    \label{fig:lambda_region}
\end{figure}

In what follows, we focus on the first resonance pole, $n=0$, and suppose that $\theta_0 < \theta < \theta_1$ to regularize the $n=0$ resonant wave function.
Note that the condition that $\theta_0<\theta<\theta_1$ provides an inequality depending on~$\lambda$, and so
\begin{align}
    - \frac{\im E^{\mathrm{R}}_0}{\re E^{\mathrm{R}}_0} < \tan 2\theta < - \frac{\im E^{\mathrm{R}}_1}{\re E^{\mathrm{R}}_1} .
\end{align}
One can find that
\begin{align}
    (\lambda < \lambda_0^- \,\lor\, \lambda_0^+ < \lambda)
    \quad \land \quad
    \lambda_1^- < \lambda < \lambda_1^+ ,
\end{align}
where
\begin{align}
    \lambda_0^\pm &= \frac{\beta^2 \hbar^2}{4m}\left[\frac{1+\tan^2 2\theta}{\tan^2 2\theta} \pm \sqrt{\left(\frac{1+\tan^2 2\theta}{\tan^2 2\theta}\right)^2 - \frac{1+\tan^2 2\theta}{\tan^2 2\theta}}\right]
    = \frac{\beta^2 \hbar^2}{4m}\frac{1 \pm \cos 2\theta}{\sin^2 2\theta} ,
    \\
    \lambda_1^\pm &= \frac{\beta^2 \hbar^2}{4m}\left[\frac{9+5\tan^2 2\theta}{\tan^2 2\theta} \pm \sqrt{\left(\frac{9+5\tan^2 2\theta}{\tan^2 2\theta}\right)^2 - \frac{9+25\tan^2 2\theta}{\tan^2 2\theta}}\right] .
\end{align}
Figure~\ref{fig:lambda_region} shows the functions~$\lambda_{0,1}^{\pm}$ with respect to~$\theta$ and indicates that $\lambda_0^+<\lambda<\lambda_1^+$ for small enough~$\theta$.

\section{Classification of wave functions and self-orthogonality}
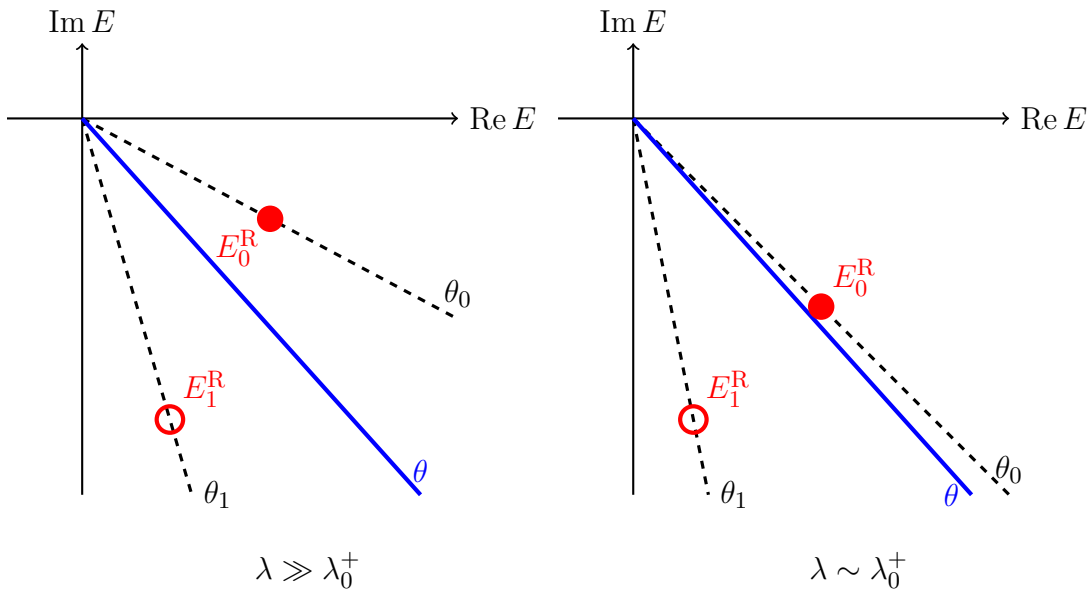
\begin{figure}[t]
    \centering
\begin{tikzpicture}
    \draw[->,thick] (-1,0) -- (5,0) node[right] {$\re E$};
    \draw[->,thick] (0,-5) -- (0,1) node[above] {$\im E$};
    \draw[dashed,very thick] (0,0) -- (5,-8/3) node[above] {$\theta_0$};
    \draw[dashed,very thick] (0,0) -- (35/24,-5) node[right] {$\theta_{1}$};
    \draw[ultra thick,blue] (0,0) -- (4.5,-5) node[above] {$\theta$};
    \fill[red] (5/2,-4/3) circle(5pt) node[below left] {$E^{\mathrm{R}}_0$};
    \draw[red,ultra thick] (7/6,-4) circle(5pt) node[above right] {$E^{\mathrm{R}}_{1}$};
    \node at (3,-6) {$\lambda\gg\lambda_0^{+}$};
\end{tikzpicture}
\begin{tikzpicture}
    \draw[->,thick] (-1,0) -- (5,0) node[right] {$\re E$};
    \draw[->,thick] (0,-5) -- (0,1) node[above] {$\im E$};
    \draw[dashed,very thick] (0,0) -- (5,-5) node[above] {$\theta_0$};
    \draw[dashed,very thick] (0,0) -- (1,-5) node[right] {$\theta_{1}$};
    \draw[ultra thick,blue] (0,0) -- (4.5,-5) node[left] {$\theta$};
    \fill[red] (5/2,-5/2) circle(5pt) node[above right] {$E^{\mathrm{R}}_0$};
    \draw[red,ultra thick] (4/5,-4) circle(5pt) node[above right] {$E^{\mathrm{R}}_{1}$};
    \node at (3,-6) {$\lambda\sim\lambda_0^{+}$};
\end{tikzpicture}
    \caption{Changing $\lambda$ and deformation of energy Riemann sheet. \textbf{Left:} For a moderate value of~$\lambda$, the resonant spectrum is completely isolated. \textbf{Right:} Near $\lambda_0^{+}$, the resonance pole is embedded into the continuum spectrum.}
    \label{fig:lambda_deformation}
\end{figure}

Let us change the parameter~$\lambda$ instead of~$\theta$ to follow the well-discussed approach to usual EPs in finite-dimensional non-Hermitian systems~\cite{Moiseyev2011NHQM}.
It is straightforward that changing $\lambda$ \revtwo{effectively deforms} $\theta_n$ and so the wave function becomes ill defined outside $\lambda_0^{+}<\lambda<\lambda_1^{+}$ (see Fig.~\ref{fig:lambda_deformation}).
Then, the parameter $\lambda$ can contact with its EP, say $\lambda_{\mathrm{bp}}$, such that $\lim_{\lambda\to\lambda_{\mathrm{bp}}} E^{\mathrm{R}}_0 = \lim_{\lambda\to\lambda_{\mathrm{bp}}} E_{\mathrm{CSM}}(\theta) \equiv E_{\mathrm{bp}}$ for a value of~$E$.

That is, precisely speaking, $\lambda_{\mathrm{bp}}$ is a root of the equation
\begin{align}
    \tan2\theta = - \frac{\im E_{\mathrm{bp}}}{\re E_{\mathrm{bp}}}
\end{align}
with the critical energy
\begin{align}
    E_{\mathrm{bp}} = \frac{\hbar^2\beta^2}{8m} \left[\sqrt{\frac{8m\lambda_{\mathrm{bp}}}{\beta^2 \hbar^2} - 1} - i \right]^2.
\end{align}
We have
\begin{align}
    \lambda_{\mathrm{bp}} = \lambda_0^+ = \frac{\beta^2 \hbar^2}{4m}\frac{1 + \cos 2\theta}{\sin^2 2\theta}    
    = \frac{\beta^2 \hbar^2}{4m}\frac{1}{1-\cos2\theta} .
\end{align}

It is important to see the behavior of the wave function around $\lambda_{\mathrm{bp}}$.
Given $\lambda$, we have the Gamow state as $\psi_{\mathrm{CSM}}(\lambda)$.
If $\theta>\theta_0(\lambda)$ then $\psi_{\mathrm{CSM}}$ is one wave function associated with the resonance with~$E^{\mathrm{R}}_0$.
Thus, the resonance is just a pseudobound state in the regularized sense by CSM.
Here, going across~$\lambda_{\mathrm{bp}}$, one may take a drastic value of~$\lambda$ such that $\theta<\theta_0(\lambda)$.
It is trivial that $\psi_{\mathrm{CSM}}$ is divergent and there exists only a resonance pole at~$n=0$.
When $\theta=\theta_0(\lambda=\lambda_{\mathrm{bp}})$, any damping factor in~$\psi_{\mathrm{CSM}}$ disappears but the plane-wave behavior at the asymptotic region is left. Actually, this is identical to the scattering solution.

\begin{figure}[t]
    \centering
\begin{tikzpicture}
    \fill[teal!10] (0,0) -- (6,-3) -- (10,-3) -- (10,0);
    \draw[teal,snake it] (5,-5/2) -- (8,0);
    \draw[teal,snake it] (3,-3/2) -- (5,0);
    \draw[teal,snake it] (1,-1/2) -- (2,0);
    \draw[->] (-1,0) -- (10,0) node[right] {$\re E$};
    \draw[->] (0,-3) -- (0,1) node[above] {$\im E$};
    \draw[red] (0,0) -- (6,-3);
    \fill (4,-2) circle(5pt) node[right] {$E_{\mathrm{bp}}$};
    \draw[ultra thick,blue] (4,-2) circle(1);
    \fill[blue] (4.5,-2.85) circle(3pt) node[below right] {$\lambda\mapsto\psi_{\mathrm{CSM}}$};
    \node at (2,-3) {Convergent};
    \node[teal] at (7,-1) {Divergent};
    \node[red] at (9,-3) {Scattering (embedded)};
\end{tikzpicture}
    \caption{The resonance wave function as a function of~$\lambda$ around~$\lambda_{\mathrm{bp}}$. Given $\lambda$, we can see the behavior of~$\psi_{\mathrm{CSM}}(\lambda)$ with~$E_{0}^{\mathrm{R}}(\lambda)$. We consider the $\lambda$ contour around~$\lambda_{\mathrm{bp}}$ as depicted by the blue circle. Below the red line, $\psi_{\mathrm{CSM}}$ is convergent and a pseudobound state. Above it, the wave function is not regularized by CSM, so $\psi_{\mathrm{CSM}}$ diverges; this region is a kind of branch plane. At intersections between the red line and blue curve, the resonance is embedded into the continuum spectrum, and its wave function is identical to the scattering solution.}
    \label{fig:lambda_contour}
\end{figure}
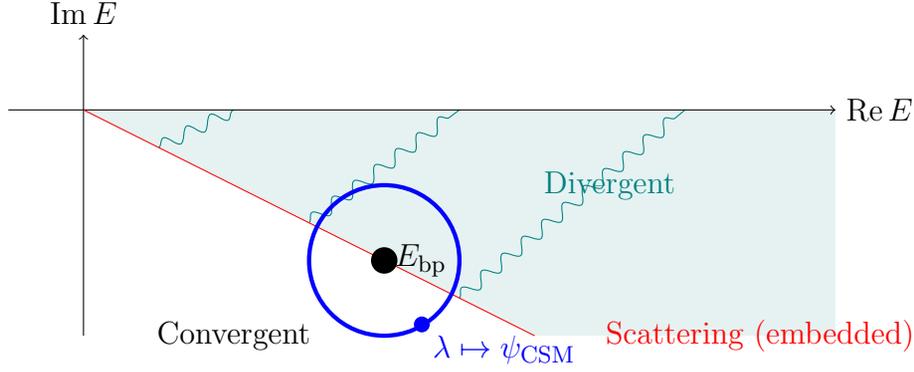

Figure~\ref{fig:lambda_contour} illustrates a contour of~$\lambda\in\mathbb{C}$ around~$\lambda_{\mathrm{bp}}$ and shows that $\psi_{\mathrm{CSM}}$ is convergent, divergent, or scattering.
The green region with the wavy lines implies the branch cut where the Gamow state should diverge; at the boundary (red line) a damping factor in the resonant wave function by CSM disappears.

\revtwo{To follow explicitly how a resonance pole approaches and coalesces with the complex-scaled continuum, we use a momentum-bin discretization of the continuum spectrum. The method and the associated notation are summarized in Appendix~\ref{sec:bin_method}, so that the present section can focus directly on the resulting branch-point structure and its physical consequences.}

The resonant wave function $\psi_{\mathrm{CSM}}^{\mathrm{R}}(\lambda)$ with $\lambda\neq\lambda_{\mathrm{bp}}$ is an independent solution of the non-Hermitian linear differential equation.
We also have the momentum-binned states $\psi_{\mathrm{CSM}}^{\mathrm{bin}}(n,\lambda)$ as the independent scattering solutions~$\psi_{\mathrm{CSM}}^{\mathrm{conti}}$ in a bin.
Originally, the linear eigenvalue problem possesses
\begin{align}
    \int \psi_{\mathrm{CSM}}^{\mathrm{R}}(\lambda) \psi_{\mathrm{CSM}}^{\mathrm{R}}(\lambda) dx &= 1 , \\
    \int \psi_{\mathrm{CSM}}^{\mathrm{bin}}(n',\lambda) \psi_{\mathrm{CSM}}^{\mathrm{bin}}(n,\lambda) dx &= \delta_{n'n}, \\
    \int \psi_{\mathrm{CSM}}^{\mathrm{R}}(\lambda) \psi_{\mathrm{CSM}}^{\mathrm{bin}}(n,\lambda) dx &= 0 .
\end{align}
\revtwo{Note that the binned states are normalized with a Kronecker delta, which makes the relevant formulas technically simpler and the branch structure easier to track explicitly. By contrast, the original continuum states are normalized with a Dirac delta, and this distribution normalization tends to obscure the algebraic steps and their physical interpretation in the present context.}

Let us consider the limit as~$\lambda\to\lambda_{\mathrm{bp}}$.
From the exact solution, we can find that
\begin{align}
    &\lim_{\lambda\to\lambda_{\mathrm{bp}}} E_{0}^{\mathrm{R}}(\lambda)
    = E_{\mathrm{bp}} ,\\
    &\lim_{\lambda\to\lambda_{\mathrm{bp}}} \psi_{\mathrm{CSM}}^{\mathrm{R}}(\lambda)
    = \psi_{\mathrm{CSM}}^{\mathrm{conti}}(k_{\mathrm{bp}},\lambda_{\mathrm{bp}}).
    \label{eq:lim_res_is_conti}
\end{align}
Thus, there exists only one general scattering solution with the identical eigenenergy.

Now, exchanging the limit and the inner product (integral over~$x$) is contradictory.
This is because
\begin{align}
    &\int \lim_{\lambda\to\lambda_{\mathrm{bp}}}\psi_{\mathrm{CSM}}^{\mathrm{R}}(\lambda) \psi_{\mathrm{CSM}}^{\mathrm{bin}}(n,\lambda) dx
    = \int \psi_{\mathrm{CSM}}^{\mathrm{conti}}(k_{\mathrm{bp}},\lambda) \psi_{\mathrm{CSM}}^{\mathrm{bin}}(n,\lambda) dx \neq 0
    &&
    \text{$\exists n$} ,\\
    &\lim_{\lambda\to\lambda_{\mathrm{bp}}}\int \psi_{\mathrm{CSM}}^{\mathrm{R}}(\lambda) \psi_{\mathrm{CSM}}^{\mathrm{bin}}(n,\lambda) dx
    = \lim_{\lambda\to\lambda_{\mathrm{bp}}} 0 = 0
    &&
    \text{$\forall n$} .
\end{align}
Therefore, the biorthogonality cannot be ensured.

To remedy this, it is natural in non-Hermitian quantum systems to introduce \textit{left} eigenvectors in addition to the \textit{right} (usual) eigenvectors.
Without exchanging the limit and integral, all right continuum states~$\psi_{\mathrm{CSM}}^{\mathrm{bin}}$ are orthogonal to a left scattering solution associated with the resonance, $\lim_{\lambda\to\lambda_{\mathrm{bp}}}\psi_{\mathrm{CSM}}^{\mathrm{R}}\sim\psi_{\mathrm{CSM}}^{\mathrm{bin}}(n_{\mathrm{bp}})$ with $\exists n_{\mathrm{bp}}$.\footnote{For simplicity, one may rewrite it as~$(\psi_{\mathrm{Left}}(n_{\mathrm{bp}}),\psi_{\mathrm{Right}}(n_{\mathrm{bp}}))=0$ for the degenerate solution~$\psi(n_{\mathrm{bp}})$ if no confusion arises.}
This is called the self-orthogonality (no biorthogonal basis).

This is a special phenomenon in non-Hermitian linear algebra so that not only the eigenvalues but also the eigenvectors are degenerate.
The Hamiltonian is nondiagonalizable but has a Jordan block.

\section{Geometric phase near resonance pole}
\revone{Here the geometric phase is defined for adiabatic transport along a closed loop in parameter space. At the exceptional point itself the eigenvectors coalesce and the Berry connection is not well defined; our results therefore pertain to loops that encircle the EP (the associated branch point), capturing the monodromy or holonomy.}

Moiseyev's ansatz of non-Hermitian energy near EP \cite{Moiseyev2011NHQM} is given by
\begin{align}
    E_{\mathrm{CSM}}^\theta(\lambda) = E_{\mathrm{bp}} + \alpha \sqrt{\lambda - \lambda_{\mathrm{bp}}} + O(\lambda),
    \label{eq:moiseyev}
\end{align}
and we define $\lambda$ as
\begin{align}
    \lambda = \lambda_{\mathrm{bp}} + R e^{i\phi} 
    \quad(R\ll1).
\end{align}
\revone{The parameter $\alpha$ in Moiseyev's ansatz is introduced to label the branch (Riemann sheet) of the analytically continued resonance energy. It is fixed once the continuation path and the continuity condition of the resonance pole are specified, and it is not a tunable parameter. The final holonomy or Berry phase is insensitive to $\alpha$ beyond this discrete sheet identification.}
Note that the square root of~$\lambda$ around~$\lambda_{\mathrm{bp}}$ comes from its branch order~$2$, that is, if $p$ wave functions coalesce into an EP then $(\lambda-\lambda_{\mathrm{bp}})^{1/p}$.\footnote{It looks hard to count the number of continuum states. Nevertheless, since there is one corresponding resonant pole/state, the rank of the projection operator to the resonance is definitely equal to one. This is the same situation as the spectrum intensity in Breit--Wigner peak of the Feshbach resonance. Thus, the associated continuum spectrum is unidentified, but one continuum state can be responsible for the EP phenomenon; it provides $p=2$.} $\alpha$ is a complex parameter.

Let us consider the geometric phase of the wave function in the CSM based on Moiseyev's ansatz. We rewrite the solution of Eq.~\eqref{eq:CSM-eq} as 
\begin{align}
    &\psi_{\mathrm{CSM}}(k,\lambda,x')= (1-\xi^2)^{-\frac{ik}{2\beta}} 
    F\left( -\frac{ik}{\beta}-s, -\frac{ik}{\beta}+s+1, -\frac{ik}{\beta}+1, \frac{1-\xi}{2} \right),
    \label{eq:CSM-sol}
\end{align}
where $k=-i\beta\kappa=\sqrt{2mE^{\theta}_{\mathrm{csm}}(\lambda)}/\hbar$ and the $k$ and $\lambda$ dependencies are written explicitly on the right-hand side. For simplicity, we discuss the geometric phase by using the asymptotic forms of the solution represented as (for instance, see Ref.~\cite{MorikawaOgawa2025PTEP})
\begin{align}
    &\psi_{\mathrm{CSM}}(k,\lambda,x')\xrightarrow{x\to\infty}4^{-\frac{ik}{2\beta}} e^{ikx'} ,
    \\  
    &\psi_{\mathrm{CSM}}(k,\lambda,x')\xrightarrow{x\to-\infty}4^{-\frac{ik}{2\beta}} e^{-ikx'} \frac{\Gamma(1-\frac{ik}{\beta})\Gamma(\frac{ik}{\beta})}{\Gamma(1+s)\Gamma(-s)} + 4^{-\frac{ik}{2\beta}} e^{ikx'}\frac{\Gamma(1-\frac{ik}{\beta})\Gamma(-\frac{ik}{\beta})}{\Gamma(-\frac{ik}{\beta}-s)\Gamma(-\frac{ik}{\beta}+s+1)}.
    \label{eq:CSM-asy}
\end{align}
The above solution corresponds to a continuum state.

To avoid some mathematical difficulties, we discretize the continuum state by using the momentum-bin method. We assume that Moiseyev's ansatz applies to a wavenumber as well if $|\lambda-\lambda_{\mathrm{bp}}|\ll1$, and we define the discretized wavenumber $k_n$ by 
\begin{align}
    k_n &= k_{\mathrm{bp}}+\alpha'_n\sqrt{\lambda-\lambda_{\mathrm{bp}}}
    \label{eq:dis-kn1}
    \\
    &= k_{\mathrm{bp}}+\alpha'_nR^{\frac{1}{2}}e^{\frac{i\phi}{2}} ,
    \label{eq:dis-kn2}
\end{align}
where $k_{\mathrm{bp}}=\sqrt{2mE_{\mathrm{bp}}}/\hbar$. Using $k_n$, we see the geometric phase as follows.

\paragraph{\textbf{Case I:} $x\to\infty$}
    The momentum-binned state at $x\to\infty$ is given by
    \begin{align}
        \psi_{\mathrm{CSM}}^{\mathrm{bin}}(n,\lambda,x') &\to \frac{1}{\sqrt{\Delta k}} \int_{k_n}^{k_{n+1}} 4^{-\frac{ik}{2\beta}} e^{ikx'}dk
        \notag\\
        &= \frac{1}{\sqrt{\Delta k}} \frac{e^{\left(-\frac{i\ln{4}}{2\beta}+ix'\right)k_{n+1}}-e^{\left(-\frac{i\ln{4}}{2\beta}+ix'\right)k_{n}}}{-\frac{i\ln{4}}{2\beta}+ix'} .
    \end{align} 
    Substituting Eq.~\eqref{eq:dis-kn1} into the above equation, we obtain
    \begin{align}
        \psi_{\mathrm{CSM}}^{\mathrm{bin}}(n,\lambda,x') &\to\frac{1}{\sqrt{\Delta k}}\frac{e^{\left(-\frac{i\ln{4}}{2\beta}+ix'\right)k_{\mathrm{bp}}}}{-\frac{i\ln{4}}{2\beta}+ix'}
        \left\{
        e^{\left(-\frac{i\ln{4}}{2\beta}+ix'\right)\alpha'_{n+1} \sqrt{\lambda-\lambda_{\mathrm{bp}}}} - e^{\left(-\frac{i\ln{4}}{2\beta}+ix'\right)\alpha'_{n}\sqrt{\lambda-\lambda_{\mathrm{bp}}}}
        \right\}.
        \label{eq:psi_inf_asymp_lambda}
    \end{align}
    Provided that the exponents of the terms in the brace brackets on the right-hand side are sufficiently small, the Taylor expansion of the asymptotic wave function is given as
    \begin{align}
        \psi_{\mathrm{CSM}}^{\mathrm{bin}}(n,\lambda,x') &\approx\frac{1}{\sqrt{\Delta k}}
        \frac{e^{\left(-\frac{i\ln{4}}{2\beta}+ix'\right)k_{\mathrm{bp}}}}{-\frac{i\ln{4}}{2\beta}+ix'} \times \left(-\frac{i\ln{4}}{2\beta}+ix'\right)
        (\alpha'_{n+1}-\alpha'_n)\sqrt{\lambda-\lambda_{\mathrm{bp}}} \notag\\
        &= \frac{1}{\sqrt{\Delta k}}
        e^{\left(-\frac{i\ln{4}}{2\beta}+ix'\right)k_{\mathrm{bp}}}
        (\alpha'_{n+1}-\alpha'_n)\sqrt{\lambda-\lambda_{\mathrm{bp}}}.
    \end{align}
    Here noting that $\Delta k=(\alpha'_{n+1}-\alpha'_n)\sqrt{\lambda-\lambda_{\mathrm{bp}}}$ and substituting Eq.~\eqref{eq:dis-kn2},
    \begin{align}
        \psi_{\mathrm{CSM}}^{\mathrm{bin}}(n,\lambda,x') &\approx e^{\left(-\frac{i\ln{4}}{2\beta}+ix'\right)k_{\mathrm{bp}}}\sqrt{\alpha'_{n+1}-\alpha'_n}R^{\frac{1}{4}}e^{i\frac{\phi}{4}} .
        \label{eq:psi_inf_asymp}
    \end{align}
    The factor $e^{i\frac{\phi}{4}}$ indicates that $\psi_{\mathrm{CSM}}^{\mathrm{bin}}(n,\lambda(\phi=0),x')|_{x\to\infty}=\psi_{\mathrm{CSM}}^{\mathrm{bin}}(n,\lambda(\phi=8\pi),x')|_{x\to\infty}$.

\paragraph{\textbf{Case II:} $x\xrightarrow{}-\infty$}
    Now, $E^{\theta}_{\mathrm{CSM}}(\lambda)$ is very close to $E_{\mathrm{bp}}$, i.e., the resonant energy. Thus, by the Siegert boundary condition \cite{Siegert:1939}, the coefficient of the second term in Eq.~\eqref{eq:CSM-asy} is negligible, and that term can be omitted. That is, the condition 
    \begin{align}
        \frac{\Gamma(1-\frac{ik}{\beta})\Gamma(-\frac{ik}{\beta})}{\Gamma(-\frac{ik}{\beta}-s)\Gamma(-\frac{ik}{\beta}+s+1)}\approx0 
    \end{align}
    implies $\frac{ik}{\beta}\approx s+1$.\footnote{Note that the complex energy of the resonance at the fourth quadrant can be given by the condition $\frac{ik}{\beta}=s+1$, while the complex energy of the anti-resonance at the first quadrant comes from $\frac{ik}{\beta}=-s$.}
    Consequently, the asymptotic form reduces to 
    \begin{align}
        \psi_{\mathrm{CSM}}(k,\lambda,x')\xrightarrow{x\to-\infty} 4^{-\frac{ik}{2\beta}} e^{-ikx'} ,
    \end{align}
    and its discrete form is 
    \begin{align}
        \psi_{\mathrm{CSM}}^{\mathrm{bin}}(n,\lambda,x') &\xrightarrow{x\to-\infty}
        \frac{1}{\sqrt{\Delta k}}\int_{k_n}^{k_{n+1}} 4^{-\frac{ik}{2\beta}} e^{-ikx'} dk
        \notag\\
        &\ \qquad = \frac{1}{\sqrt{\Delta k}} \frac{e^{\left(-\frac{i\ln{4}}{2\beta}-ix'\right)k_{n+1}}-e^{\left(-\frac{i\ln{4}}{2\beta}-ix'\right)k_{n}}}{-\frac{i\ln{4}}{2\beta}-ix'} .
    \end{align}
    This is essentially the same as \textbf{Case I}, and a similar calculation yields the factor $e^{i\frac{\phi}{4}}$.\footnote{Alternatively, if the coefficient of the first term in Eq.~\eqref{eq:CSM-asy} is retained exactly, Euler’s reflection formula leads to the asymptotic form
    \begin{align}
        4^{-\frac{ik}{2\beta}} e^{-ikx'} \frac{i\sin(\pi s)}{\sinh(\frac{\pi k}{\beta})} .
    \end{align}
    Using the formula
    \begin{align}
        \frac{1}{\sinh(\frac{\pi k}{\beta})} = 2 \sum_{n=0}^{\infty} e^{-(2n+1)\frac{\pi k}{\beta}} \qquad \left(\re \frac{\pi k}{\beta} >0\right) ,
    \end{align}
    the momentum-binned state is modified as
    \begin{align}
        \psi_{\mathrm{CSM}}^{\mathrm{bin}}(n,\lambda,x') &\xrightarrow{x\to-\infty} 2\frac{\sin(\pi s)}{\sqrt{\Delta k}} \sum_{n=0}^{\infty} \int_{k_n}^{k_{n+1}} e^{\left(-\frac{i\ln{4}}{2\beta}-ix'-\frac{(2n+1)\pi}{\beta}\right)k}dk
        \notag\\
        &\ \qquad = 2\frac{\sin(\pi s)}{\sqrt{\Delta k}} \sum_{n=0}^{\infty} \frac{e^{\left(-\frac{i\ln{4}}{2\beta}-ix'-\frac{(2n+1)\pi}{\beta}\right)k_{n+1}}-e^{\left(-\frac{i\ln{4}}{2\beta}-ix'-\frac{(2n+1)\pi}{\beta}\right)k_{n}}}{-\frac{i\ln{4}}{2\beta}-ix'-\frac{(2n+1)\pi}{\beta}} .
    \end{align}
    Thus, the same calculation as in \textbf{Case I} gives $e^{i\frac{\phi}{4}}$ as an overall factor.}
    Thus, the same periodicity condition, $\psi_{\mathrm{CSM}}^{\mathrm{bin}}(n,\lambda(\phi=0),x')|_{x\to-\infty}=\psi_{\mathrm{CSM}}^{\mathrm{bin}}(n,\lambda(\phi=8\pi),x')|_{x\to-\infty}$, is satisfied.

\revone{I}ncorporating \textbf{Case I} and \textbf{Case II}, when we consider the contour depicted in Fig.~\ref{fig:lambda_contour}, that is, encircling the critical energy~$E_{\mathrm{bp}}$ by changing~$\lambda$ (or $\phi$), the wave function should have the $8\pi$ periodicity (branch order is $4$). This periodicity or branch structure is quite universal in non-Hermitian quantum mechanics with a $2\times 2$ Jordan block. It is, however, much more nontrivial and remarkable that the additional branch structure arises due to the discretized momentum-bin size~$\Delta k$, and so it essentially depends on the continuous (unbounded) parameter~$k$ after removing the cutoff scale.
\revone{Here the additional branch structure is traced to the square-root normalization factor~$1/\sqrt{\Delta k}$
in Eq.~\eqref{eq:psi_inf_asymp_lambda} that arises from momentum-bin discretization. While this introduces an apparent dependence on the bookkeeping parameter $\alpha_n$
[cf. Eq.~\eqref{eq:psi_inf_asymp}], the associated branch choice does not affect the final result: it is precisely this square-root branch that guarantees that the discretized formulation reproduces the Berry phase.}

\begin{figure}[t]
    \centering
\begin{tikzpicture}
    \fill[teal!10] (0,0) -- (6,-3) -- (8,-3) -- (8,0);
    \draw[teal,snake it] (5,-5/2) -- (8,0);
    \draw[teal,snake it] (3,-3/2) -- (5,0);
    \draw[teal,snake it] (1,-1/2) -- (2,0);
    \draw[->] (0,0) -- (8,0) node[right] {$\re E$};
    \draw[->] (0,-3) -- (0,1) node[above] {$\im E$};
    \draw[red] (0,0) -- (6,-3);
    \fill (4,-2) circle(5pt) node[right] {$E_{\mathrm{bp}}$};
    \draw[ultra thick,blue] (4,-2) circle(1);
    \fill[blue] (4.9,-2.45) circle(3pt) node[left] {$\lambda^{+}$};
    \fill[blue] (3.1,-1.55) circle(3pt) node[right] {$\lambda^{-}$};
    \draw[ultra thick,red,dotted,->] (5,-2.8) -- (5.5,-2);
    \draw[ultra thick,red,dotted,->] (3,-1) -- (2.5,-1.8);
    \node at (2,-3) {A: $\psi_{\mathrm{CSM}}^{\mathrm{R}}$};
    \node[teal] at (8,-1) {B: $\psi_{\mathrm{CSM}}^{\mathrm{conti}}$};
    \node[red] at (6,-3.5) {$\lim_{\lambda\nearrow}\psi_{\mathrm{CSM}}^{\mathrm{R}}=\psi_{\mathrm{CSM}}^{\mathrm{conti}}$};
\end{tikzpicture}
\begin{tikzpicture}
    \draw[very thick] (0,0) circle(0.1) node[right] {$E_{\mathrm{bp}}$};
    \fill[blue] (-2,0) circle(0.1) node[below left] {$\lambda^{-}$};
    \fill[blue] (2,0) circle(0.1) node[below right] {$\lambda^{+}$};
    \draw[ultra thick,teal,->] (2,0) arc(0:180:2);
    \node[left,teal] at (0,2.2) {$\phi\in[0,2\pi]$};
    \node[teal] at (0,0.5) {$E_-=E_{\mathrm{bp}}-\alpha R^{\frac{1}{2}}e^{\frac{i\phi}{2}}$};
    \draw[ultra thick,->] (-2,0) arc(180:360:2);
    \node[right] at (0,-2.2) {$\phi\in[0,2\pi]$};
    \node at (0,-0.6) {$E_+=E_{\mathrm{bp}}+\alpha R^{\frac{1}{2}}e^{\frac{i\phi}{2}}$};
    \draw[ultra thick,red,dashed,->] (-0.2,0) -- (-1.95,0) node[left] {$\arg\alpha$};
\end{tikzpicture}
    \caption{The resonance and continuum wave functions in the complex energy plane. The contour around~$E_{\mathrm{bp}}$ is depicted by the blue circle. Below the red line, i.e., in the region A, $\psi_{\mathrm{CSM}}^{\mathrm{R}}$ is well defined. Otherwise, in the region B, we focus on~$\psi_{\mathrm{CSM}}^{\mathrm{conti}}$. At intersections between the red line and blue curve, say $\lambda^{+}$ and~$\lambda^{-}$, the resonance is identical to the continuum state~$\psi_{\mathrm{CSM}}^{\mathrm{conti}}$ along the limit of~$\lambda$ as the left and/or right eigenfunction. We set the argument of~$\alpha$ as the right panel so that Moiseyev's ansatz~\eqref{eq:moiseyev} becomes $E(\lambda^{-})$ when $\phi=0$.}
    \label{fig:psi_on_lambda}
\end{figure}
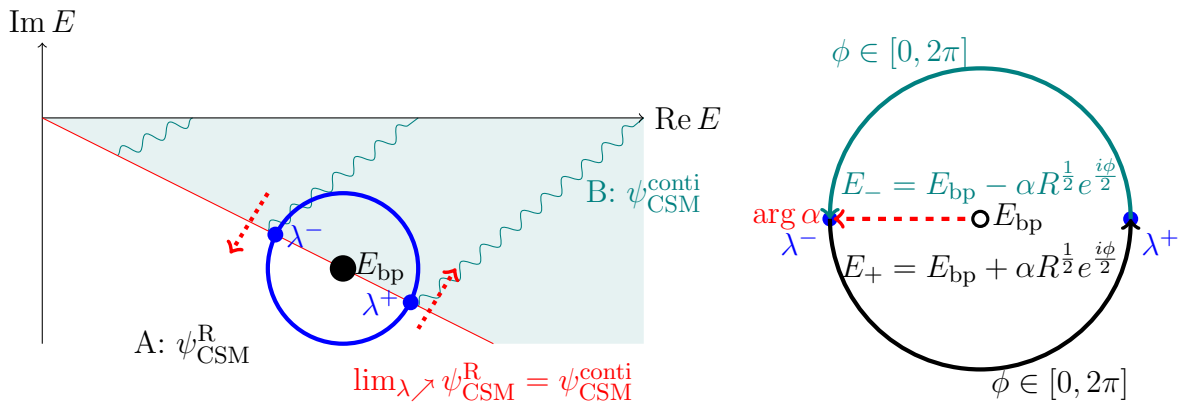

To gain more physical insight, let us see what has happened by the $\phi$ rotation.
Figure~\ref{fig:psi_on_lambda} shows how to encircle $E_{\mathrm{bp}}$ by the current rotation.
The region A denotes the area where $\psi_{\mathrm{CSM}}^{\mathrm{R}}(\lambda(\phi))$ is normalizable, while the region B implies that any resonance is not regularized by CSM.
At the boundary between A and B, the resonant state becomes the continuum one as Eq.~\eqref{eq:lim_res_is_conti}.
Also we set $\arg\alpha$ as the right panel in Fig.~\ref{fig:psi_on_lambda} so that Moiseyev's ansatz~\eqref{eq:moiseyev} becomes $E(\lambda^{-})$ when $\phi=0$.
In effect, now, we can decompose the complex-energy regions as
\begin{align}
    E_{\pm} = E_{\mathrm{bp}} \pm \alpha R^{\frac{1}{2}} e^{\frac{i\phi}{2}} 
    \qquad
    \phi\in[0,2\pi].
\end{align}
Here, the resonant state possesses its complex eigenenergy by~$E_+$ in this convention.
Suppose that $\psi_{\mathrm{CSM}}^{\mathrm{R}}(\lambda(0|_A))$ and $\psi_{\mathrm{CSM}}^{\mathrm{conti}}(\lambda(0|_B))$ are taken as the exact solution~\eqref{eq:CSM-sol} (i.e., \textit{their overall phase conform to that of the explicit expression}).

First, we start from an intersection point in the region A defined by~$\phi=0$, so $\psi_{\mathrm{CSM}}^{\mathrm{R}}(\lambda(0|_A))=\psi_{\mathrm{CSM}}^{\mathrm{R}}(\lambda^{-})$.
Second, we deform $\lambda$ near the vicinity of another intersection, that is, $\lambda(2\pi|_A)=\lambda^{+}$. Note that the overall factor~$e^{\frac{i\phi}{4}}$ exists in each wave function, by which the dilated resonant wave function differs from the continuum state. We thus use the \revone{continuity} condition as
\begin{align}
    \psi_{\mathrm{CSM}}^{\mathrm{R}}(\lambda(2\pi|_A))=e^{\frac{i\pi}{2}}\psi_{\mathrm{CSM}}^{\mathrm{conti}}(\lambda(0|_B))=i\psi_{\mathrm{CSM}}^{\mathrm{conti}}(\lambda(0|_B)) .
\end{align}
Next, by deformation in the region B, we have
\begin{align}
    \psi_{\mathrm{CSM}}^{\mathrm{conti}}(\lambda(2\pi|_B))=e^{\frac{i\pi}{2}}\psi_{\mathrm{CSM}}^{\mathrm{R}}(\lambda(0|_A))=i\psi_{\mathrm{CSM}}^{\mathrm{R}}(\lambda(0|_A)) .
\end{align}
If the complex energy rotates once, the wave function acquires a minus sign:
\begin{align}
    \psi_{\mathrm{CSM}}^{\mathrm{R}}(\lambda(4\pi|_A))
    = - \psi_{\mathrm{CSM}}^{\mathrm{R}}(\lambda(0|_A)) .
\end{align}
This is the reason why $\phi$ has the $4\times2\pi$ periodicity.
We have observed the nontrivial Berry phase induced by the resonance pole as an EP.

\section{Chern characteristic and monodromy renormalization}
Finally, we consider a simple extension: the Chern-type characteristic of the quantum resonance pole as an EP in the parameter space.
We \revone{show} that it is associated with a resonance branch point in the $\lambda$-parameter space, and a renormalization viewpoint on the monodromy (holonomy) when continuum normalization becomes distributional.

\subsection{Biorthogonal connection and bin-regularized holonomy}
We first define an Abelian connection for a (generically) non-Hermitian eigenproblem.
It is natural to define the Chern connection by
\begin{align}
    \mathcal{A}^\theta(\lambda)
    \equiv i \int \psi_{\mathrm{CSM}}^{\mathrm{bi(Left)}} \frac{\partial}{\partial\lambda} \psi_{\mathrm{CSM}}^{\mathrm{(Right)}} dx
    = i ( \psi(\lambda), \partial_\lambda \psi(\lambda)).
\end{align}
Here, $\psi^{\mathrm{(Left)}}$ denotes a left eigenvector, and $\psi^{\mathrm{(Right)}}$ is an associated right eigenvector (biorthogonal partner); those are corresponding to~$\psi_{\mathrm{CSM}}^{\mathrm{R}}$ or~$\psi_{\mathrm{CSM}}^{\mathrm{conti,bin}}$ in an appropriate way.
We then define the holonomy around the branch point $\lambda_{\mathrm{bp}}$ as
\begin{align}
    \gamma &\equiv \oint_{\mathrm{bp}} \mathcal{A}^\theta(\lambda) d\lambda,
    &
    c_1 &\equiv \frac{1}{2\pi}\gamma,
    \label{eq:holonomy_chern_character}
\end{align}
where $c_1$ is a Chern-type characteristic (first Chern number when the underlying line bundle is globally well defined).

We next evaluate $\gamma$ within the momentum-bin regularization and focus on the most singular contribution in $\lambda-\lambda_{\mathrm{bp}}$.
For simplicity, suppose that from the asymptotic relation in Eq.~\eqref{eq:psi_inf_asymp}
\begin{align}
    \frac{\partial}{\partial\lambda}\psi_{\mathrm{CSM}}^{\mathrm{bin}}(n,\lambda,x')
    \approx \frac{1}{4}\frac{1}{\lambda-\lambda_{\mathrm{bp}}}\psi_{\mathrm{CSM}}^{\mathrm{bin}}(n,\lambda,x') .
\end{align}
In this bin-level estimate, the left or right distinction affects only nonsingular gauge choices and does not modify the holonomy; hence we suppress it for notational simplicity.\footnote{More precisely, the holonomy $\gamma$ is invariant under $\psi\to e^{i\chi(\lambda)}\psi$ and the corresponding transformation of $\psi^{\mathrm{bi}}$, so the leading singular piece is sufficient for $\gamma$.}
Then, a nontrivial Abelian phase of holonomy around the resonance pole is obtained if $n=n_{\mathrm{bp}}$ by
\begin{align}
    \gamma(n_{\mathrm{bp}}) &\equiv \oint_{\mathrm{bp}} \mathcal{A}^\theta(n_{\mathrm{bp}},\lambda) d\lambda \\
    &= i \oint_{\mathrm{bp}} \int \psi_{\mathrm{CSM}}^{\mathrm{bin(Left)}}(n_{\mathrm{bp}},\lambda,x') \frac{\partial}{\partial\lambda} \psi_{\mathrm{CSM}}^{\mathrm{bin(Right)}}(n_{\mathrm{bp}},\lambda,x') dx d\lambda \notag\\
    &= i \oint_{\mathrm{bp}} \frac{1}{4}\frac{1}{\lambda-\lambda_{\mathrm{bp}}} d\lambda
    = \frac{\pi}{2} .
\end{align}
The four-rotations invariance is reflected in $e^{i\gamma}\in\mathbb{Z}_4$. The Chern characteristic suffers from the fractionality of its naive Chern number; the corresponding characteristic $c_1=\gamma/2\pi=1/4$ is fractional in this bin-regularized sense.
That is, it indicates that the resonance eigenstate under CSM does not define a globally rigid (single-valued) line bundle over the punctured
$\lambda$-plane without passing to a branched cover.

\subsection{Continuum limit: Two sources of ill definedness}
So far, the momentum-bin method and the exact solution of the inverted Rosen--Morse potential have enabled us to provide the explicit estimates in all of our observations.
Averaging continuum states within a momentum bin is a technique in finite volume regularization.
The continuum eigenvectors are, however, defined intrinsically by a distribution-valued construction (rigged Hilbert space, von~Neumann ring with affiliated operators, semi-infinite measure, and so on).
\revone{Even in the formal continuum spectrum, the physical picture remains intact and is not particularly sensitive to the choice of geometric path, so we do not confine the discussion to geometric paths alone.}
On the other hand, the definition of~$\mathcal{A}(\lambda)$ becomes quite subtle due to the plausible normalization factor as~$\delta(0)$ for the continuum states.
Hence, in this section, the vector-bundle structure induced by~$\lambda$ is not straightforwardly applied to the continuum case.

Here is a short summary:
The problem is divided into (A) inner products of the continuum states and (B) the number of those.
(A) Continuum eigenvectors are $\delta$-function normalized, and inner products produce regulator-dependent singular factors such as $\delta(0)$ (or its derivatives). This affects the very meaning of $\mathcal{A}^\theta(\lambda)$ unless a specific regularization (finite volume box, wave-packet, test-function, etc.) is fixed.
(B) Even after choosing a regulator, the set of continuum states is uncountable and the associated bundlelike structure over parameter space is not automatically inherited straightforwardly.
  
To make the appearance of (A) explicit, in general, we have an asymptotic form of the scattering/resonant solution near the resonance pole
\begin{align}
    \psi_{\mathrm{CSM}}^{\mathrm{as}}(k,\lambda,x') =
    \begin{cases}
        \frac{1}{\sqrt{2\pi}} e^{ikx'} & \re x'\gg0, \\
        \frac{1}{\sqrt{2\pi}} e^{-ikx'} & \re x'\ll0,
    \end{cases}
\end{align}
where $k=k_{\mathrm{bp}}+\alpha'\sqrt{\lambda-\lambda_{\mathrm{bp}}}$ and this is $\delta$-function normalized.
With attention to the self-orthogonality, we obtain
\begin{align}
    \mathcal{A}^\theta(\lambda)
    &= i \lim_{k'\to k,\lambda'\to\lambda}
    \int \psi_{\mathrm{CSM}}^{\mathrm{as(Left)}}(k',\lambda',x') \frac{\partial}{\partial\lambda} \psi_{\mathrm{CSM}}^{\mathrm{as(Right)}}(k,\lambda,x') dx \\
    &= i \lim_{k'\to k,\lambda'\to\lambda}
    \frac{\partial k}{\partial\lambda} \frac{\partial}{\partial k}
    \int \psi_{\mathrm{CSM}}^{\mathrm{as(Left)}}(k',\lambda',x')  \psi_{\mathrm{CSM}}^{\mathrm{as(Right)}}(k,\lambda,x') dx\\
    &= i \left(\frac{\alpha'}{2\sqrt{\lambda-\lambda_{\mathrm{bp}}}} + \frac{\partial\alpha'}{\partial\lambda} \sqrt{\lambda-\lambda_{\mathrm{bp}}}\right) \delta'(0),
    \label{eq:berry_as_delta}
\end{align}
where we have assumed that $\alpha'=\alpha'(\lambda)$ to be $\lambda$ dependent.
Here $\delta'(0)$ is a shorthand for the regulator-dependent singular factor arising from differentiating the continuum overlap; its precise meaning depends on the chosen regularization scheme (e.g., box normalization or wave-packet regularization).
The naive connection requires renormalization/regularization data in the
continuum.

\subsection{Monodromy renormalization: Fixing the physical holonomy}
$\gamma$ takes measurements of the monodromy around the EP.
Now, we implement the monodromy renormalization method in Ref.~\cite{Morikawa:2025ezl} as a way to separate physical content
from regulator dependence.
Let us remove the microscopic circle around~$\lambda_{\mathrm{bp}}$ with the radius~$R$.
Then, the physical holonomy being kept $\pi/2$, i.e., $\gamma^{\mathrm{phys}}=\pi/2$, the \textit{bare} holonomy depends on~$R$ as
\begin{align}
    \gamma^{\mathrm{phys}} = \frac{\gamma^{\mathrm{bare}}(R)}{Z_{\gamma}(R)} .
\end{align}
$Z_\gamma(R)$ plays the role of a renormalization factor that absorbs regulator dependence [including distributional normalization factors such as $\delta(0)$ or $\delta'(0)$].
Still, $\gamma^{\mathrm{bare}}$ can be divergent by~$\delta(0)$, but only the renormalized monodromy is observable.

Notice that it appears that $\alpha$ ($\alpha'$) is a measure of the number of continuum states along the $E$ ($k$) contour.
In fact, the structure of $k=k_{\mathrm{bp}}+\alpha' R^{\frac{1}{2}}e^{\frac{i\phi}{2}}$ suggests that the coefficient $\alpha'$ controls how many continuum states effectively contribute along the $k$ contour (equivalently, along the energy contour).
Motivated by this, we introduce a redefinition (running) of $\alpha'(\lambda)$ such that the singular factor $\delta'(0)$ is
absorbed into $\alpha'$.
We impose the renormalization-group-type condition
\begin{align}
    \alpha' + 2\frac{\partial\alpha'}{\partial\lambda}(\lambda-\lambda_{\mathrm{bp}}) = \frac{\alpha_0}{2\sqrt{\lambda-\lambda_{\mathrm{bp}}}\delta'(0)} \label{eq:renor_group}
\end{align}
for finite~$\alpha_0$, and the solution is given by
\begin{align}
    \alpha'(\lambda) = \frac{\alpha_0}{4\delta'(0)} \frac{\ln(\lambda-\lambda_{\mathrm{bp}})}{\sqrt{\lambda-\lambda_{\mathrm{bp}}}} + \frac{\text{(const.)}}{\sqrt{\lambda-\lambda_{\mathrm{bp}}}} .
\end{align}
Under this redefinition, the $\delta$ function from the connection can be absorbed by the \textit{renormalization} of~$\alpha'$.\footnote{$\delta'(0)$ should not be interpreted as a literal number: the connection is meaningful only after specifying a regularization and/or
considering smeared states (test functions), and in the present discussion its dependence is absorbed into renormalized quantities.
For instance, in a box of length $L$ one has $\delta(0)\sim L/2\pi$ (and derivatives thereof depend on boundary
conditions), while in a wave-packet regularization $\delta$ is replaced by a narrow function of width $\epsilon$.
The renormalization group equation should therefore be read as a compact way to track this regulator dependence.}

Then, the resulting holonomy becomes finite and proportional to $\alpha_0$:
\begin{align}
    \gamma^{\mathrm{bare}}(R)=\frac{\pi}{2}\alpha_0.
\end{align}
The physical monodromy is kept fixed while the regulator dependence is absorbed into $Z_{\gamma}(R)=\alpha_0$ and the running parameter $\alpha'(\lambda)$.
Eventually, while~$\gamma^{\mathrm{phys}}=\pi/2$ in this way, the bin-level result and the continuum viewpoint can be unified within a renormalization framework, providing a consistent physical picture of monodromy around a resonance branch point in the presence of continuum normalization singularities.


\section{Conclusions}
\revone{Exceptional points (EPs) are commonly discussed as defective spectral singularities of non-Hermitian (more generally, non-self-adjoint) operators, and their topology is encoded in the branch structure of eigenvalues and the associated exchange of eigenvectors.
In this work, we revisited this EP topology from the viewpoint of \revthree{quantum scattering theory} 
and quantum resonances in an infinite-dimensional setting.
Using the complex-scaling method (CSM), we regularized Gamow (Siegert) states and exposed resonance poles as discrete complex eigenvalues separated from the complex-rotated continuum.}

To treat resonant and scattering states on the same footing, we introduced a momentum-bin discretization for the complex-scaled continuum spectrum.
This construction provides a practical spectral representation of the nonunitary dilated Hamiltonian in which the resonance state and a (separable) set of complex-scaled momentum-binned states can be continuously tracked as the parameter $\lambda$ is varied.
Within this framework, we clarified how an EP arises as the coalescence of a resonance with the discretized continuum. A branch point structure associated with a resonance pole embedded in the continuum spectrum has played a central role.
In particular, we highlighted the self-orthogonality at the EP: the biorthogonal normalization breaks down, and the resonant and continuum eigenstates are degenerate.
\revtwo{In the discretized representation, the effective spectral description of the resonance-continuum sector becomes defective at the branch point.} \revtwo{This defectiveness is reflected in the emergence of a Jordan-block structure familiar from non-Hermitian linear algebra.}

We then studied the geometric phase acquired by the complex-scaled resonant wave function under a deformation of $\lambda$ around the branch point $\lambda_{\mathrm{bp}}$.
Now, the wave function must become intrinsically multivalued.
As we explicitly observed, after a $4\pi$ rotation in the $\lambda$~plane the resonant wave function acquires a minus sign. This observation implies a $4\times 2\pi$ periodicity and a nontrivial Berry phase and holonomy class induced by the resonance pole.
This provides a direct bridge between the EP Berry phase familiar in finite-dimensional non-Hermitian models and the analytic structure of resonance poles in scattering theory.

There are several natural extensions from the present formulation.
It is interesting to analyze the Stokes graph/topology of the complex-scaled differential equation under the resonance-continuum coalescence.
The real-time evolution of unstable states in this setting is very attractive \revtwo{for future study},
\revtwo{especially} near EP-like physics.

\paragraph{Operational Meaning and Observable Signatures}
\revone{We finally comment on the operational meaning of the parameter loop and possible observable signatures in a scattering setting. In practice, an adiabatic variation of a control parameter $\lambda$ should be understood in a quasi-static sense: one considers a slowly time-dependent potential $V[x;\lambda(t)]$ and probes scattering at a fixed incoming energy $E$ while $\lambda(t)$ is cycled. A natural criterion for quasistatic driving is that the driving timescale $T_{\mathrm{drive}}$ is much longer than the intrinsic response time of the scattering process. The latter can be characterized by the Wigner time delay $\tau_W(E;\lambda)$, defined by
\begin{equation}
\tau_W(E;\lambda)=\frac{d}{dE}\,\delta(E;\lambda),
\end{equation}
where $\delta(E;\lambda)$ is the scattering phase shift [equivalently, $\tau_W=d\,\arg S(E;\lambda)/dE$ in a single-channel setting].\footnote{%
\revone{In multichannel scattering, a natural generalization is given by the Smith time-delay matrix $Q(E;\lambda)=-i\,S^\dagger(E;\lambda)\,\frac{d}{dE}S(E;\lambda)$, whose eigenvalues define the eigendelays; one may also work with eigenphases obtained from the eigenvalues of $S(E;\lambda)$.}}
The quasistatic condition then reads
\begin{equation}
T_{\mathrm{drive}}\gg \tau_W(E;\lambda)
\quad \text{for all $\lambda$ along the loop.}
\end{equation}}

\revone{Under such a slow closed cycle in parameter space that encloses the relevant branch point, the holonomy discussed in this work is expected to be encoded in the analytic structure of the scattering matrix $S(E;\lambda)$. Consequently, the cycle may leave an imprint on observable scattering characteristics, such as the accumulated change in the phase shift $\delta(E;\lambda)$ (or, equivalently, the phase of $S$), and possibly on resonance line shapes when the loop traverses a parameter region where the pole structure of $S$ changes appreciably. Establishing a detailed quantitative connection between the holonomy of the complex-scaled resonance eigenstates and experimentally accessible scattering observables requires a time-dependent scattering analysis and is beyond the scope of the present work; we leave it for future investigation.}

\section*{Acknowledgements}
We are grateful to the restaurant Ajisawa for their hospitality.
This work was partially supported by Japan Society for the Promotion of Science (JSPS) Grant-in-Aid for Scientific Research Grant Numbers JP25K17402 (O.M.), JP21H04975 (S.\ Ogawa), and JP25KJ1954 (S.\ Onoda).
O.M.\ acknowledges the RIKEN Special Postdoctoral Researcher Program
and RIKEN FY2025 Incentive Research Projects.

\appendix

\section{Complex-scaled continuum states in momentum-bin method}\label{sec:bin_method}
\revone{Since treating the continuum spectrum entails substantial technical difficulties, we introduce a discretization scheme; in particular, applying it also to the CSM makes subsequent computations considerably more tractable.
In this \revtwo{Appendix}, we briefly review the momentum-bin method and fix our notation, mainly to make the paper self-contained.}

\subsection{Hermitian quantum mechanics}
It is complicated to treat continuum states because they are intrinsically $\delta$-function normalized. Let $\phi(k,x)$ denote a continuum state with a momentum $k$, whose orthogonality is determined by
\begin{align}
    \int \phi^*(k',x) \phi(k,x) dx = \delta(k'-k) .
\end{align}
We therefore discretize the continuum states using the momentum-bin method, in which the resulting bin states provide an accurate representation of the continuum states in realistic nuclear reaction calculations~\cite{Yahiro2012CDCC}. In this method, a continuum state is averaged over the width of the momentum bin $\Delta k$.

It is then natural to define a discretized state by
\begin{align}
    \Hat{\phi}_n (x) = \frac{1}{\sqrt{\Delta k}} \int_{k_n}^{k_{n+1}}\phi(k,x) dk,
    \qquad
    k_{n+1} - k_n = \Delta k,\,
    n\in\mathbb{Z}.
\end{align}
The orthogonality of the discretized states is shown as
\begin{align}
    \int\Hat{\phi}_{n'}^*(x) \Hat{\phi}_n (x) dx &= \frac{1}{\Delta k} \int_{k'_{n'}}^{k'_{n'+1}} \int_{k_n}^{k_{n+1}} \delta(k'-k) dk' dk
    \notag\\
    &= \delta_{n'n} .
\end{align}
\revone{One may use the bracket notation as
\begin{equation}
    \langle\Hat\phi_{n'},\Hat\phi_n\rangle=\delta_{n'n}.
\end{equation}}
Moreover, a Hamiltonian $h$ is diagonalized with regard to~$\Hat{\phi}_n (x)$ as follow:
in scattering problem, from the Schr\"odinger equation
\begin{align}
    h \phi(k,x) = \frac{\hbar^2k^2}{2m} \phi(k,x) ,
\end{align}
we find that
\begin{align}
    \int\Hat{\phi}_{n'}^*(x) h \Hat{\phi}_n (x) dx &= \frac{1}{\Delta k} \int_{k'_{n'}}^{k'_{n'+1}} \int_{k_n}^{k_{n+1}} \frac{\hbar^2 k^2}{2m} \delta(k'-k) dk' dk
    \notag\\
    &= \Hat{\epsilon}_n \delta_{n'n} ,
\end{align}
where
\begin{align}
    \Hat{\epsilon}_n \equiv \frac{1}{\Delta k} \int_{k_n}^{k_{n+1}} \frac{\hbar^2 k^2}{2m}dk .
\end{align}
The relation between $\Hat{\phi}_n(x)$ and $\phi(k,x)$ is now represented as
\begin{align}
    \int \phi^*(k,x) \Hat{\phi}_n(x) dx &= 
    \begin{cases}
        \frac{1}{\sqrt{\Delta k}} & k_n \le k \le k_{n+1} ,\\
        \ \ \ 0 & \text{otherwise} .
    \end{cases}
\end{align}

As a simple example, the momentum-bin discretization of a plane wave
\begin{align}
    \frac{1}{\sqrt{\Delta k}} \int_{k_n}^{k_{n+1}} e^{ikx} dk = \frac{1}{\sqrt{\Delta k}} \frac{e^{ik_{n}x}-e^{ik_{n+1}x}}{ix}
\end{align}
falls off as $1/x$ for large $x$.

\subsection{Complex-scaled Hamiltonian}
The momentum-bin method is well-established in Hermitian quantum mechanics. In this section, we extend it to non-Hermitian quantum mechanics. Let us consider the complex-scaled continuum state $\phi^{\theta}(k,x)$, which is a solution of the complex-scaled Hamiltonian $h^{\theta}$.
\revone{It is natural to consider that} the discretization of the state is given by
\begin{align}
    \Hat{\phi}_n^{\theta} (x) &= \frac{1}{\sqrt{\Delta k}} \int_{k_n}^{k_{n+1}} \phi^{\theta}(k,x) dk .
\end{align}
The biorthogonal state of $\phi^{\theta}(k,x)$ is $\phi^{*\theta}(k,x)$ (see Chap.\ 6.1 in Ref.~\cite{Moiseyev2011NHQM}). Thus, the biorthogonal state of $\Hat{\phi}_n^{\theta} (x)$ is represented as
\begin{align}
    \Hat{\phi}_n^{\mathrm{bi},\theta} (x) &\equiv \frac{1}{\sqrt{\Delta k}} \int_{k_n}^{k_{n+1}} \phi^{*\theta}(k,x) dk .
\end{align}
Then, the orthogonality of the scaled and discretized state is shown as
\begin{align}
    \int \Hat{\phi}_{n'}^{\mathrm{bi},\theta} (x) \Hat{\phi}_n^{\theta} (x) dx &= \frac{1}{\Delta k} \int_{k'_{n'}}^{k'_{n'+1}} \int_{k_n}^{k_{n+1}} \left\{ \int \phi^{*\theta}(k',x) \phi^{\theta}(k,x) dx \right\} dk'dk
    \notag\\
    &=\frac{1}{\Delta k} \int_{k'_{n'}}^{k'_{n'+1}} \int_{k_n}^{k_{n+1}} \delta(k'-k) dk'dk
    \notag\\
    &=\delta_{n'n}.
\end{align}
\revone{The corresponding bracket notation is given as
\begin{equation}
    (\Hat\phi_{n'},\Hat\phi_n)=\delta_{n'n}, 
\end{equation}
which is the bilinear inner product.}

\bibliographystyle{utphys}
\bibliography{ref,ref_ewkb,ref_res,ref_nonhermitian}
\end{document}